\documentstyle[aps,eqsecnum,preprint]{revtex}
\begin{document}
\draft
\preprint{\vbox{\hbox{SNUTP 94-54} \hbox{hep-ph/yymmddd} \hbox{June 1994} }}
\title{Instanton Contribution to \\
        $B \rightarrow X_{s} \gamma $ Decay}
\author{Junegone Chay\footnote{\tt chay@kupt.korea.ac.kr}
      and Soo-Jong Rey\footnote{\tt sjrey@phyb.snu.ac.kr}}
\address{Physics Department, Korea University, Seoul 136-701 Korea${}^*$ \\
         Physics Department and Center for Theoretical Physics \\
         Seoul National University, Seoul 151-752 Korea${}^\dagger$ }
\date{June 10, 1994}
\maketitle

\begin{abstract}
We study instanton effects on $B \rightarrow X_s\gamma$ decay using
the heavy quark effective field theory
and the operator product expansion. In the dilute gas
approximation the effect is negligibly small.
This is in contrast to the result of the instanton effect in inclusive
semileptonic $b\rightarrow u$ decay but similar to the inclusive
hadronic $\tau$ decay. We discuss the similarities and differences of
the $B \rightarrow X_s\gamma$ decay compared to inclusive hadronic
$\tau$ decay and semileptonic $B$ decay.
\end{abstract}
\narrowtext
\section {Introduction}
Rare decays of hadrons containing a single $b$ quark are extensively
studied as a  sensitive probe to new physics beyond the standard model.
At the electroweak scale the new physics introduces extra contribution
to the coefficient functions of local operators in the low-energy
effective theory \cite{inamilim,grinsteinwise}.
Radiative $b \rightarrow s \gamma$ decay is of such type which is
induced by penguin diagrams with virtual heavy quarks in the
loop. Recent CLEO collaboration has reported the branching ratio
of $B \rightarrow K^* \gamma$ as $(4.5 \pm 1.9 \pm 0.9) \%$
\cite{cleo}. However theoretical estimates of the branching
ratio are model dependent. So far all the theoretical estimates of
the exclusive channels have been based on either the
quark model \cite{quarkmodel} or the QCD sum rules \cite{sumrule}.
The main uncertainty lies in evaluating the
matrix element of the penguin operator between hadronic states.
Therefore it is expected to be more reliable to consider the inclusive
radiative $B$ decays in which all the final states containing a single
$s$ quark are  summed over.

In calculating the inclusive decay rate we follow the approach
of Chay et.al. \cite{chayetal} to use the heavy quark effective
field theory (HQEFT) and the operator product expansion (OPE).
Bigi et.al. \cite{bigietal}, Neubert \cite{neubert} and Falk et.al.
\cite{falketal} have performed the analysis and have found that
the leading contribution is the parton-model result. They have
systematically obtained corrections coming from the hadronic
matrix elements of higher dimensional operators.
The size of the corrections is about 5\% of the parton-model result.
There are other sources of corrections. Perturbative QCD
corrections have been studied by Ali and Grueb \cite{aligrueb}.

One also expects additional nonperturbative QCD corrections.
In the case of $B \rightarrow X_u e \overline \nu_e$ decay
we have calculated the
instanton contribution using the dilute gas approximation ~\cite
{chayrey} and have found that the instanton contribution to the total
decay rate is large. In order to make the perturbation theory
reliable, it is necessary to introduce a smearing prescription.
In this paper we study the nonperturbative effects induced by
instantons in $B \rightarrow X_s \gamma$ decay.
We find that the instanton contribution to the total decay rate in
this case is small.
The striking difference of the instanton contribution in both
decays may be surprising. We account for this difference
by comparing the kinematics in the two cases.

This paper is organized as follows. In Section 2 using the OPE and
HQEFT we formulate the decay rate from the effective Hamiltonian
relevant to the $B \rightarrow X_s\gamma$ decay.
We relate the rate to the forward Compton scattering amplitude and
discuss the analytic structure of the amplitude. In Section 3
we study the instanton contribution to the decay rate as a calculable
example of nonperturbative corrections to the decay rate. We calculate
explicitly the instanton contribution at leading order.
In Section 4 we discuss the kinematics of the $b \rightarrow
s\gamma$ decay and compare it to that of the $b \rightarrow u e
\overline \nu_e$ and the inclusive hadronic $\tau$ decays.
We find that the kinematics of the radiative
$B$-meson decay is different from that of the semileptonic $B$ decay
but similar to that of the inclusive hadronic $\tau$ decay.
We explain how the kinematics affects the size of the instanton
contribution in these cases.
In Section 5 we summarize our results and give a conclusion.

\section{Operator Product Expansion}
In the standard model the effective Hamiltonian for $b\rightarrow s$
decay consists of various operators. They are generated after
integrating out the $W$ bosons, Higgs particle and heavy fermions at
the weak scale.
The effective Hamiltonian describing $b \rightarrow s$ decay can be
written as
\begin{equation}
H_{\rm eff} =  -{4 G_F \over \sqrt 2} V_{tb}V_{ts}^*
\sum_{j} c_j(\mu) {\cal O}_j(\mu).
\end{equation}
For inclusive $b \rightarrow s \gamma$ decay only
the operator ${\cal O}_7$ contributes, which is
\begin{equation}
{\cal O}_7 = {e \over 16 \pi^2}
(m_b \overline s_{L} \sigma^{\mu \nu}b_{R}
+m_s \overline s_{R} \sigma^{\mu \nu}b_{L} ) F_{\mu \nu}.
\label{operators}
\end{equation}
The effective Hamiltonian
for $b\rightarrow s\gamma$ decay at $\mu =m_b$ is written as
\begin{equation}
H_{\rm eff} = -{4 G_F \over \sqrt 2} V_{tb} V^*_{ts} c_7 (m_b) {\cal
O}_7 (m_b)
\label{effH}
\end{equation}
where the Wilson coefficient $c_7 (m_b)$ is scaled down from $\mu =
M_W$  to $\mu = m_b$ using the renormalization group equation.

The radiative $B$-meson decay rate is given by
\begin{equation}
d \Gamma = {4 G_F^2 \over m_b}|V_{tb} V^*_{ts}|^2 |c_7(m_b)|^2
     \sum_{X_s,\lambda} (2 \pi)^4 \delta^{(4)}(p_B - p_X - k)
|\langle X_s \gamma |{\cal O}_7 | B \rangle |^2 d ({\rm PS}),
     \label{partonrate}
     \end{equation}
where the sum includes all the possible hadronic final states $X_s$
containing an $s$ quark and the photon polarizations $\lambda$.
$p_B^\mu = m_b v^\mu$ and $k^\mu$ are the four momenta of the $B$
meson and the photon respectively. The phase space is written as
\begin{equation}
d ({\rm PS}) = \int {d^3 k \over (2 \pi)^3} {1 \over 2 E_\gamma}
= {m_b^2 \over 16 \pi^2}\int dy \, y
\label{phasespace}
\end{equation}
where we have rescaled the photon energy as $ y = 2 E_\gamma / m_b$,
$0 \le y \le 1$.
Using the explicit form of the operator ${\cal O}_7$ and summing over
photon polarizations we get
\begin{eqnarray}
d \Gamma &=& \frac{\alpha G_F^2 m_b}{256\pi^5} |V_{tb}V^*_{ts}|^2
|c_7(m_b)|^2 y \, dy \nonumber \\
&\times& \sum_{X_s, \lambda} (2 \pi)^4 \delta^{(4)}
(p_B - p_X - k) |\langle X_s, \gamma | (m_b {\overline s}_L + m_s
{\overline s}_R)\sigma^{\mu\nu}F_{\mu\nu} b| B\rangle|^2
\label{decay}
\end{eqnarray}
The matrix element-squared in Eq.~(\ref{decay}) can be factorized as
\begin{eqnarray}
&\sum_{X_s}& (2 \pi)^4 \delta^{(4)}
(p_B - p_X - k) \langle B | {\overline b}\sigma^{\mu\nu} (m_b s_L +
m_s s_R)|X_s\rangle \nonumber \\
&\times&\langle X_s| (m_b {\overline s}_L + m_s
{\overline s}_R)\sigma^{\alpha\beta} b| B\rangle \nonumber
\\
&\times& \sum_{\lambda}\langle 0|F_{\mu\nu}|\gamma ;\lambda,k\rangle
\langle \gamma ;\lambda, k | F_{\alpha\beta}|0\rangle.
\label{factorize}
\end{eqnarray}
The matrix elements of $F_{\mu\nu}$ are written as
$\langle 0 | F_{\mu \nu} | \gamma; k, \lambda \rangle =
i (k_\mu \epsilon_{\nu}^{(\lambda)}(k)- k_\nu
\epsilon_{\mu}^{(\lambda)}(k))$.
Summing over photon polarizations,
$\sum_{\lambda}\epsilon_{\mu}^{(\lambda)}(k)
\epsilon_{\nu}^{(\lambda)}(k) =-g_{\mu \nu}$, we have
\begin{equation}
\sum_\lambda  \langle 0 | F_{\mu \nu} | \gamma ; k, \lambda \rangle
\langle  \gamma ; k, \lambda |  F_{\alpha \beta}| 0 \rangle
= (k_\nu k_\alpha g_{\mu \beta} - k_\mu k_\alpha g_{\nu \beta}
+k_\mu k_\beta g_{\nu \alpha} - k_\nu k_\beta g_{\mu \alpha}).
\label{photon}
\end{equation}
The inclusive decay rate is related to the quantity $W$ defined
by
\begin{eqnarray}
W &=& (2\pi)^3\sum_{X_s} \delta^{(4)} ( p_B - p_X - k) \nonumber \\
&\times&\langle B | {\overline b} [ \not{k} , \gamma_\nu ](m_b s_L +
m_s s_R) | X_s \rangle \langle X_s | (m_b {\overline s}_L + m_s
{\overline s}_R) [\not{k}, \gamma^\nu] b | B \rangle .
\end{eqnarray}
$W$ depends on the photon momentum $k^{\mu}$.
In the complex $y = 2 v \cdot k / m_b$ plane with $k^2$ fixed,
$W$ is related to the
discontinuity of the forward Compton scattering amplitude $T(v \cdot k)$
across a physical cut as
\begin{equation}
W = 2\ \mbox{Im}\ T.
\end{equation}

For $B$ decays it is appropriate to use the HQEFT in which the full
QCD $b$ field is expressed in terms of $b_v$ for $b$-quark velocity
$v$. The $b_v$ field is defined by
the heavy $b$ field $ b_v(x) = e^{i m_b v \cdot x} b (x)$ which
satisfies $v \hskip-0.22cm / b_v = b_v$.
We can apply the techniques of the OPE to
expand $T$ in terms of matrix elements of local operators involving
$b_v$ fields
\begin{eqnarray}
T(v \cdot k) &=& -i\int d^4 x e^{i (m_b v - k)\cdot x}
\langle B | T \{ J^\mu (x) \, J_\mu^\dagger (0) \} | B\rangle
\nonumber \\
     &=& \sum_{n,v}C_n^v(k)\langle B | {\cal O}_v^{(n)} |B \rangle,
\label{Compton}
\end{eqnarray}
where
\begin{equation}
J^{\mu} = {\overline b}_v[\not{k}, \gamma^{\mu}](m_b s_L + m_s s_R).
\end{equation}
In Eq.~(\ref{Compton}) ${\cal O}_v^{(n)}$ are local operators
involving ${\overline b}_v b_v$ bilinears. The coefficient functions
$C_n^v$ depend explicitly on $k$. Short-distance physics is contained
in the coefficient functions while the matrix elements describe
long-distance physics.

In order to evaluate the decay rate it is necessary to examine the
analytic structure of $T$. In the complex $y$ plane the
cut relevant to the decay is located on the real axis $y\leq 1-r$
where $r = m_s^2/m_b^2$. The cut extends down below $y=0$, which is
the boundary of the physical region for the decay process (in which the
photon has a lightlike momentum). We will refer to the region along the
real axis for
\begin{equation}
0\leq y \leq 1-r
\label{physcut}
\end{equation}
as the ``physical cut''. There is also another cut
\begin{equation}
y\geq (2+\sqrt{r})^2 -1
\end{equation}
corresponding to other physical processes. These cuts in the complex
plane are shown in Fig.~1. The discontinuity of $T$ along this physical
cut yields $W$ for $b\rightarrow s\gamma$ decay. The physically allowed
region for $y$ is $0\leq y \leq 1-r$. We can introduce the
cutoff of the photon's energy such that $y_c\leq y \leq 1-r$. This
cutoff is necessary experimentally since it is difficult to observe
a soft photon.

As $y\rightarrow 1 - r$, that is, at the right end of the
physical cut, the invariant mass of the hadronic system approaches
$m_s$ and we enter the region where resonances dominate. We expect large
perturbative QCD corrections in this region. However we are able to
reliably compute suitable averages of physical quantities by
integrating around the physical cut along the contour $C$ shown in
Fig.~1. The contour integral along $C$ and $C^{\prime}$ contains no
singularities, so the integral vanishes. Thus the contribution along
the cut directly related to the discontinuity is minus the
contribution from the contour integral along the contour $C$
which stays far away from the resonance region.

The distance between the right end of the physical cut and the left
end of the other cut is 2 while the maximum length of the physical
cut is 1 for $y_c=0$ and $r=0$. Therefore we can reliably calculate the
contour integral along $C$ without encountering the other cut.
$T$ can be expanded in powers of $\alpha_s$ and $1/m_b$.
The leading term in $1/m_b$ is given by
\begin{equation}
T = \frac{-i}{2}\int d^4 x e^{i {\cal Q} \cdot x} \langle B |
{\overline b}_v(x)
[\not{k} , \gamma_\mu] (1+\gamma_5)S_F(x)(1-\gamma_5) [ \not{k},
\gamma^\mu] b_v (0) | B \rangle ,
\label{ope}
\end{equation}
where ${\cal Q} = m_b v -k$ is the momentum of the $s$ quark and
$S_F(x)$ denotes the $s$ quark propagator. In momentum space
the leading term of Eq.~(\ref{ope}) in $\alpha_s$ is proportional to
\begin{equation}
\frac{{\cal Q} \hskip-0.23cm / + \not{k_b}+m_s}{({\cal Q}+k_b)^2-m_s^2},
\end{equation}
where $k_b$ is the residual momentum of the $b_v$ field of order
$\Lambda$.

For $\Lambda^2 \ll {\cal Q}^2 \ll m_b^2$ it is sufficient
to keep the terms at leading order in $1/m_b$ only and expand
Eq.~(\ref{ope}) in powers of $k_b/|{\cal Q}|$. This generates
the coefficient functions of the local operators in which $k_b$ is
replaced by the covariant derivatives acting on the $b_v$ fields.
The leading term independent of $k_b$ is given by
\begin{equation}
T = 16 m_b^3(1+r) \frac{k \cdot {\cal Q} k \cdot v}{{\cal
Q}^2-m_s^2},
\label{parton}
\end{equation}
which reproduces the parton-model result.
Here we have used the normalization
$\langle B | \overline b_v \gamma^\mu b_v | B \rangle = 2 m_b v^\mu$.
Eq.~(\ref{parton}) can be expressed in terms of $y$ as
\begin{equation}
T_0 = - 4 m_b^4 (1+r){ y^2 \over { y - 1 + r}}.
\label{parton0}
\end{equation}
The imaginary part gives rise to a delta function at the endpoint
\begin{equation}
W = 2 \ \mbox{Im}\ T = 8 \pi m_b^4 (1+r) y^2\delta (1 - y-r).
\end{equation}
We finally obtain
\begin{equation}
\frac{d\Gamma}{dy} = \Gamma_0 \delta (1 - y-r),
\end{equation}
where
\begin{equation}
\Gamma_0 = \frac{\alpha m_b^5 G_F^2}{32\pi^4} |V_{tb}V^*_{ts}|^2
|c_7(m_b)|^2 (1+r)(1-r)^3
\end{equation}
is the total decay rate in the parton model.

\section{Instanton Contribution}
We now compute the instanton effect to the decay rate using the
dilute gas approximation.
We emphasize that we have not attempted to calculate other
nonperturbative effects such as multi-instantons or renormalons.
It is possible that they give bigger contributions than what we compute.
Neverthless our estimate may still characterize a typical size of
the nonperturbative effects.

In estimating the contribution we analytically continue
the amplitude into the Euclidean region of the kinematic variables where
${\cal Q}^2$ is large enough to use the OPE reliably.
In the background of an instanton ($+$ anti-instanton) of size $\rho$
and instanton orientation $U$ located at the origin, the Euclidean
fermion propagator may be expanded in small fermion mass
as~\cite{andreigross}
\begin{eqnarray}
\displaystyle
S_\pm (x, &y;& \rho_\pm; U_\pm)  =
- {1 \over m} \psi_0 (x) \psi^\dagger_0 (y)
+ S^{(1)}_\pm (x, y; \rho_\pm; U_\pm) \nonumber \\
&+& m \int d^4 w S^{(1)}_\pm (x, w; \rho_\pm; U_\pm)
S^{(1)}_\pm (w, y; \rho_\pm; U_\pm)
\nonumber \\
&+& {\cal O}(m^2),
\label{propexp}
\end{eqnarray}
where $\pm$ denotes instanton, anti-instanton.
$\psi_0$ is the fermion zero mode eigenfunction  and
$S^{(1)}_\pm = \sum_{E>0} {1 \over E} \Psi_{E \pm} (x)
\Psi^\dagger_{E \pm} (y)$ is the Green function of fermion
nonzero modes.

In evaluating the forward Compton scattering amplitude
$T$, Eq.~(\ref{ope}), we use the propagator in Eq.~(\ref{propexp})
instead of the free propagator. $T$ can be written as
\begin{equation}
T  = T_0  + T_{\rm inst},
\label{tdecomp}
\end{equation}
where the first term is the parton-model amplitude.
The second term is the amplitude due to instantons of all
orientations $U$, position $z$ and size $\rho$.
After averaging over instanton orientations $T_{\rm inst}$
is given by
\widetext
\begin{eqnarray}
T_{\rm inst}   &=& -i \frac{m_b^2}{2} \int \! d^4 \Delta \,
e^{i {\cal Q} \cdot \Delta } \sum_{a = \pm} \!\! \int
\! d^4 z_a \, d \rho_a D(\rho_a)\nonumber \\
&\times& \langle B | \overline b_v (x) [\not{k}, \gamma^\mu]
(1+\gamma_5) \{ {\cal S}_a (X, Y; \rho_a) -S_0 (\Delta)\}
[\not{k} , \gamma_\mu]
(1 - \gamma_5) b_v (y) | B \rangle,
\label{instt}
\end{eqnarray}
\narrowtext
\noindent where $D(\rho)$ is the instanton density,
$\Delta = x - y$, $X = x - z_a$ and $Y = y - z_a$.
In Eq.~(\ref{instt}) ${\cal S}_\pm (X, Y; \rho_\pm)$ is the
fermion propagator averaged over instanton (anti-instanton)
orientations centered at $z$ and $S_0$ is the free fermion
propagator. Using the ${\overline {\rm MS}}$ scheme with $n_f$
flavors of light fermions, $D(\rho)$ is given by
\begin{equation}
\displaystyle
D(\rho)  =  K \, \Lambda^5 \, (\rho \Lambda)^{6 + {n_f \over 3}}
\, \biggl( \ln{1 \over \rho^2 \Lambda^2}
\biggr)^{45-5 n_f \over 33 - 2 n_f},
\label{density}
\end{equation}
where
\begin{eqnarray}
\displaystyle
K  & = & \biggl( \prod_i {\hat m_i \over \Lambda} \biggr) \,
2^{12 n_f \over 33 - 2n_f} \, \biggl({33 - 2n_f \over 12} \biggr)^6
 \nonumber \\
& \times & {2 \over \pi^2} \,\exp\bigl[{1 \over 2} - \alpha (1)
+ 2 (n_f -1) \alpha ({1 \over 2}) \bigr]
\label{ddensity}
\end{eqnarray}
in which the $\beta$ function at two loops and the running mass
at one loop are used and $\hat m_i$ are the
renormalization-invariant quark masses.
In Eq.~(\ref{ddensity}) $\alpha(1) = 0.443307$ and $\alpha(1/2) =
0.145873$. From now on we replace the logarithmic term in $D(\rho)$ by
its value for $\rho = 1/|{\cal Q}|$.  Corrections to this replacement
are negligible since they are logarithmically suppressed.

In Eq.~(\ref{instt}), due to the chirality structure, the nonzero-mode
Green function ${\cal S}^{(1)}_\pm$ is the leading contribution in the
${\cal O}(m_b^2)$ and ${\cal O}(m_s)^2$ terms while the zero-mode Green
function contributes to ${\cal O}(m_b m_s)$ term. However the
${\cal O}(m_b m_s)$ term is proportional to $k^2$. This vanishes since the
outgoing photon is on the mass shell.
Therefore the leading instanton contribution comes only from the
nonzero-mode Green function.
For simplicity we neglect the $m_s^2 / m_b^2$ subleading term.

The difference of the single-instanton propagator and the free
propagator in momentum space after integrating over the instanton
orientation, position and size is given by ~\cite{chayrey}
\begin{equation}
iN_{\rm inst} \frac{{\cal Q} \hskip-0.23cm /}{({\cal Q}^2)^7}
\biggl(\ln \frac{{\cal Q}^2}{\Lambda^2} \biggr)^{10/9} ,
\label{insprop}
\end{equation}
where
\begin{equation}
N_{\rm inst}= K \frac{2^{10}}{35} \pi^2 \Gamma^2(6) \Lambda^{12}.
\end{equation}
The exponent of the logarithm is almost unity and we replace it by one.
We expect that this does not change the result much.
Following the same calculation as in Ref.~\cite{chayrey},
we find the instanton contribution to $T$ is
\begin{equation}
T_{\rm inst} (v \cdot k) = 16 m_b^3 N_{\rm inst} \frac{k\cdot {\cal Q}
k\cdot v}{({\cal Q}^2)^7}
\ln {{\cal Q}^2 \over \Lambda^2}.
\label{instt2}
\end{equation}
For simplicity we put $r=0$ in Eq.~(\ref{instt2}). The inclusion of
nonzero $r$ is straightforward. In terms of the rescaled kinematic
variables we find
\begin{equation}
T_{\rm inst} = \frac{4N_{\rm inst}}{m_b^8} {y^2 \over (1 - y)^7}
\ln \bigl[{m_b^2  \over \Lambda^2} (1 - y)\bigr].
\end{equation}

Instanton effects on the $B \rightarrow X_s\gamma$ decay depend on the
momentum cutoff of the final-state photon. The cutoff to remove soft
photons should be introduced in experiments since it is difficult to
isolate a soft photon from the background coming
from the subsequent decay of, say, $K^* \rightarrow K + \gamma$.
Experimentally the photon spectrum from $B \rightarrow X_s\gamma$ is
concentrated in the region $2.2\ {\rm GeV} \lesssim E_{\gamma}
\lesssim 2.5
\ {\rm GeV}$. In CLEO the lower cut is taken at $ E_\gamma \sim 2.2
$GeV ~\cite{cleo}. Thus we restrict the
region of contour integral to $y_c \le y \le 1$ where $y_c$ is
the experimental cutoff for the scaled photon energy.
Then the instanton decay rate is
\begin{equation}
\Gamma_{\rm inst} (y_c) =  \Gamma_0
\ \frac{N_{\rm inst}}{60m_b^{12}} \
\biggl[\frac{10}{(1-y_c)^6}-\frac{36}{(1-y_c)^5}+ \frac{45}{(1-y_c)^4}-
\frac{20}{(1-y_c)^3}\biggr].
\label{rate}
\end{equation}
For $n_f = 3$, the numerical value of $K$ is approximately
$126 \ {\hat m_u \hat m_d \hat m_s / \Lambda^3}$.
Using the renormalization-group invariant
quark masses $\hat m_u = 8.2 \pm 1.5$ Mev, $\hat m_d = 14.4 \pm 1.5 $ MeV
and $\hat m_s = 288 \pm 48 $ MeV ~\cite{quarkmass},
Eq.~(\ref{rate}) is written as
\begin{equation}
\frac{\Gamma_{\rm inst} (y_c)}{\Gamma_0}
= \biggl(\frac{6.67 \mbox{GeV}}{m_b} \biggr)^3
\ \biggl({\Lambda \over m_b} \biggr)^{9} \
\biggl[\frac{10}{(1-y_c)^6}-\frac{36}{(1-y_c)^5}+ \frac{45}{(1-y_c)^4}-
\frac{20}{(1-y_c)^3}\biggr].
\end{equation}
The ratio $\Gamma_{\rm inst}(y_c)/\Gamma_0$ is shown in Fig.~2 as a
function of the cutoff $y_c$ for different values of
$\Lambda = 350, 400, 450 $ MeV with $m_b = 5$ GeV.
For $y_c \approx 0.82$ that CLEO has chosen, the instanton correction
is $(1.37 \sim 13.1) \times 10^{-4}$ as $\Lambda$ varies from $350$ MeV
to $450$ MeV. Therefore the instanton correction is negligibly small
compared to other corrections.
Note that the instanton contribution is much smaller as the cutoff
$y_c$ decreases.
On the other hand, as Fig.2 shows, the contribution is appreciable
at $y_c \gtrsim 0.92$. In this region we need a smearing to make the
perturbation theory valid.
However this cutoff is too large to be significant experimentally
since there is a very small fraction of the rate in this energy window.

The $\alpha_s$-correction to the decay rate also becomes large as $y_c$
approaches $1$. Ali and Grueb ~\cite{aligrueb} have examined the
correction in detail and have concluded that
the $\alpha_s$-correction has to be exponentiated near the endpoint
of the photon spectrum.
Numerically they have found that the exponentiation is necessary for
$y_c \gtrsim 0.85$.
Combined with their result our analysis indicates that there is
a large theoretical uncertainty for the cutoff $y_c \gtrsim 0.88$.
Therefore in order to compare experiments with theory in a
model-independent way the cutoff $y_c$ has to be chosen below 0.88.

We also obtain the instanton contribution to the differential decay
rate as
\begin{equation}
\frac{d \Gamma_{\rm inst}}{d y} = \Gamma_0 \frac{N_{\rm inst}}{m_b^{12}}
{y^3 \over (1 - y)^7}.
\end{equation}
Numerically we find
\begin{equation}
\frac{1}{\Gamma_0} \frac{d \Gamma_{\rm inst}}{d y}
= \biggl(\frac{26.1\ \mbox{GeV}}{m_b} \biggr)^3
\bigl(\frac{\Lambda}{m_b} \bigr)^9 {y^3 \over (1 - y)^7}.
\label{diffratio}
\end{equation}
The ratio in Eq.~(\ref{diffratio}) is shown in Fig.3 for different values
of $\Lambda = 350, 400, 450 $ GeV with $m_b = 5$ GeV.
The ratio also increases sharply near the endpoint of the photon spectrum.
Above $y \approx 0.87$ the instanton correction is appreciable
and necessitates a smearing in this region.

\section{Discussion on Kinematics}
We have found that the instanton effect is small in the total decay
rate of $B \rightarrow X_s \gamma$ as long as $y_c$ is small enough.
This is in contrast to the case of $B \rightarrow X_u e \overline
\nu_e$ decay. This apparently big difference can be understood from
kinematics.

As explained in Section 2 we have obtained the averaged total decay rate
by deforming the integration contour (from $C'$ to $C$ in Fig.1).
If the deformed contour is chosen sufficiently far away from the
resonance region we can calculate the decay rate reliably using the
perturbation theory. It depends on the details of the kinematics whether
such contour deformation is possible.

In $B\rightarrow X_s\gamma$ decay the radius of the deformed contour $C$
is fixed at $1 - y_c$. This radius represents the off-shell
invariant mass-squared of the final hadrons.
Note that this radius is independent of kinematic variables.
As long as $1 - y_c \gg \Lambda^2 / m_b^2$ the averaged decay rate
can be calculated reliably at the scale $m_b^2 (1 - y_c)$.
The instanton contribution is evaluated at this scale.
As calculated in Section 3, the contribution is ${\cal O}(10^{-4})$.

In $B \rightarrow X_u e \overline \nu_e$ decay the radius of the
deformed contour is given by $z = (y - \hat q^2)(1 - y) / y$ where
$ y = 2 E_e /m_b$ and $\hat q^2 = (k_e + k_\nu)^2 / m_b^2$ ~\cite{chayrey}.
The off-shell invariant mass-squared of the final hadrons is equal
to $m_b^2 z$. As long as $z \gg \Lambda^2 / m_b^2$ we can calculate
the averaged decay rate using perturbation theory. However unlike the
$B\rightarrow X_s\gamma$ case the radius $z$ depends on how the momentum
transferred to the leptonic system is distributed between the electron
and the anti-neutrino.
In particular $z$ vanishes at the boundaries of the phase space.
Therefore the instanton contribution evaluated at the scale $m_b^2 z$
grows rapidly near the boundaries $y = 1$ and $y= \hat q^2$.
This means that it is impossible to avoid the resonance region unless we
introduce a model-dependent cut near the boundaries of the phase space.

Note that the dependence of the off-shell invariant mass-squared of
the final hadrons on the kinematic variables arises only when there
are more than one ``nonhadronic'' particle in the final state.
In this respect it is interesting to compare our result with the
inclusive hadronic $\tau$ decay $\tau \rightarrow \nu_{\tau} +X$
{}~\cite{braaten}. In this case the neutrino plays the same role as the
photon in the $B\rightarrow X_s\gamma$ decay as far as the kinematics
is concerned. The maximum invariant mass-squared $s = q^2$ of the final
hadrons is fixed at $m_{\tau}^2$ independent of the neutrino energy.
Since this is much larger than $\Lambda^2$ the total inclusive decay
rate can be calculated reliably using the deformed contour.
While the specific kinematic variables under consideration are different,
the analytic structure and the fact that the maximal invariant
mass-squared of the final hadrons is independent of other kinematic
variables are similar in both $B \rightarrow X_s \gamma$ and inclusive
hadronic $\tau$ decays.

Indeed the instanton contributions are small in both cases. Nason and
Porrati~\cite{porrati} have obtained the instanton contribution to the
ratio of the hadronic to the leptonic width $R_{\tau}$ which is given
by
\begin{equation}
\frac{R_{\tau}^{\rm inst}}{R_{\tau}^0} =
\biggl(\frac{3.64\Lambda}{m_{\tau}}\biggr)^9 \frac{{\hat m}_u {\hat m}_d
{\hat m}_s}{m_{\tau}^3}.
\end{equation}
Our result for the instanton contribution to the decay rate of
$B \rightarrow X_s \gamma$ is written as
\begin{equation}
\frac{\Gamma_{\rm inst}}{\Gamma_0} =
\biggl(\frac{5.91\Lambda}{m_b}\biggr)^9 \frac{{\hat m}_u {\hat m}_d
{\hat m}_s}{m_b^3}
\end{equation}
for $y_c = 0$. The fact that both have the same mass dependence and
similar numerical factors confirms our expectation.

By the same argument we expect the instanton contribution to the
$B \rightarrow X_s e^+ e^-$ decay rate is similar to that of the
$B \rightarrow X_u e \overline \nu_e$ because kinematics
and the analytic structure of the forward Compton scattering amplitudes
are the same.

\section{Conclusion}
In this paper we have calculated the instanton effect on the decay
rate of $B \rightarrow X_s\gamma$. Unlike $B \rightarrow
X_u e \overline \nu_e$ decay we have found that the instanton correction
is negligibly small. We have also observed that the kinematics
and the analytic structure in this case is quite similar to the
inclusive hadronic $\tau$ decay, in which the instanton
correction is also small.

It is interesting to note that the contribution of nonperturbative
effects like instantons depends on kinematics. It depends on
whether kinematics allows to choose a contour in the complex plane of
the relevant variables to avoid the resonance region and to use
perturbation theory reliably.
In the inclusive semileptonic $B$ decay it is impossible to choose
a deformed contour to avoid the resonance region.
Therefore we need a smearing prescription which limits the theoretical
prediction in a model-independent way.

When we probe new physics from the radiative inclusive $B$ decay,
it is therefore sufficient to examine only those corrections coming
from the hadronic matrix elements and the radiative QCD corrections.
The latter part is $\sim 30\%$ of the parton-model result.
On the other hand since the QCD coupling constant is sensitive to the
choice of the renormalization scale $\mu$ at one-loop order,
it is necessary to set the scale $\mu$ precisely.
This requires QCD corrections beyond one-loop.
Partial attempt along this direction has been made recently by
Buras et.al. \cite {burasetal}.

\section*{Acknowledgements}
This work was supported in part by KOSEF-SRC program (SJR, JGC), KOSEF
Grant 941-0200-022-2 (JGC), Ministry of Education BSRI-94-2418 (SJR) and
BSRI-94-2408 (JGC), KRF-Nondirected Research Grant`93 (SJR).

\begin{figure}
\caption{Anaytic structure of $T$ in the complex $y$ plane. The point
$P$ at $y=y_c$ is the lower cutoff of the photon energy.
The contour integral
along $C^{\prime}$ is minus the contour integral along $C$.}
\end{figure}
\begin{figure}
\caption{The ratio $\Gamma_{\rm inst} (y_c) / \Gamma_0 $ for $\Lambda =
350, 400, 450$ MeV with $m_b = 5$ GeV.}
\end{figure}
\begin{figure}
\caption{The ratio $(1 / \Gamma_0) d \Gamma_{\rm inst} / d y$ for
$\Lambda = 350, 400, 450$ MeV with $m_b = 5$ GeV.}
\end{figure}
\end{document}